\documentclass[preprint,twocolumn]{aastex6}
\usepackage{amsmath,amssymb}
\usepackage{graphicx} 
\usepackage{natbib} 
\usepackage{aas_macros} 
\slugcomment{}
\shortauthors{Hirano et al.}
\shorttitle{RM Measurements for TRAPPIST-1}
\begin{document}
\title{
Evidence for Spin-orbit Alignment in the TRAPPIST-1 System
}
\author{
Teruyuki Hirano\altaffilmark{1},
Eric Gaidos\altaffilmark{2},
Joshua N.\ Winn\altaffilmark{3},
Fei Dai\altaffilmark{4},
Akihiko Fukui\altaffilmark{5,6},
Masayuki Kuzuhara\altaffilmark{7,8},
Takayuki Kotani\altaffilmark{7,8,9},
Motohide Tamura\altaffilmark{7,8,10},
Maria Hjorth\altaffilmark{11},
Simon Albrecht\altaffilmark{11},
Daniel Huber\altaffilmark{12},
Emeline Bolmont\altaffilmark{13},
Hiroki Harakawa\altaffilmark{14},
Klaus Hodapp\altaffilmark{15},
Masato Ishizuka\altaffilmark{10},
Shane Jacobson\altaffilmark{15},
Mihoko Konishi\altaffilmark{16},
Tomoyuki Kudo\altaffilmark{14},
Takashi Kurokawa\altaffilmark{8,17},
Jun Nishikawa\altaffilmark{7,8},
Masashi Omiya\altaffilmark{7,8},
Takuma Serizawa\altaffilmark{17},
Akitoshi Ueda\altaffilmark{8},
Lauren M. Weiss\altaffilmark{12}
}
\altaffiltext{1}{Department of Earth and Planetary Sciences, Tokyo Institute of Technology,
2-12-1 Ookayama, Meguro-ku, Tokyo 152-8551, Japan}
\email{hirano@geo.titech.ac.jp}
\altaffiltext{2}{Department of Earth Sciences, University of Hawaii at M\={a}noa, Honolulu, HI 96822, USA}
\altaffiltext{3}{Department of Astrophysical Sciences, Princeton University, 4 Ivy Lane, Princeton, NJ 08544, USA}
\altaffiltext{4}{Division of Geological and Planetary Sciences, California Institute of Technology,1200 East California Blvd, Pasadena, CA 91125, USA}
\altaffiltext{5}{Department of Earth and Planetary Science, Graduate School of Science, The University of Tokyo, 7-3-1 Hongo, Bunkyo-ku, Tokyo 113-0033, Japan}
\altaffiltext{6}{Instituto de Astrof\'isica de Canarias, V\'ia L\'actea s/n, E-38205 La Laguna, Tenerife, Spain}
\altaffiltext{7}{Astrobiology Center, NINS, 2-21-1 Osawa, Mitaka, Tokyo 181-8588, Japan}
\altaffiltext{8}{National Astronomical Observatory of Japan, NINS, 2-21-1 Osawa, Mitaka, Tokyo 181-8588, Japan}
\altaffiltext{9}{Department of Astronomy, School of Science, The Graduate University for Advanced Studies (SOKENDAI), 2-21-1 Osawa, Mitaka, Tokyo, Japan}
\altaffiltext{10}{Department of Astronomy, Graduate School of Science, The University of Tokyo, 7-3-1 Hongo, Bunkyo-ku, Tokyo 113-0033, Japan}
\altaffiltext{11}{Stellar Astrophysics Centre, Department of Physics and Astronomy, Aarhus University, Ny Munkegade 120, DK-8000 Aarhus C, Denmark}
\altaffiltext{12}{Institute for Astronomy, University of Hawai'i, 2680 Woodlawn Drive, Honolulu, HI 96822, USA}
\altaffiltext{13}{Observatoire Astronomique de l'Universit\'e de Gen\`eve, Universit\'e de Gen\`eve, CH-1290 Versoix, Switzerland}
\altaffiltext{14}{Subaru Telescope, 650 N. Aohoku Place, Hilo, HI 96720, USA}
\altaffiltext{15}{University of Hawaii, Institute for Astronomy, 640 N. Aohoku Place, Hilo, HI 96720, USA}
\altaffiltext{16}{Faculty of Science and Technology, Oita University, 700 Dannoharu, Oita 870-1192, Japan}
\altaffiltext{17}{Tokyo University of Agriculture and Technology, 3-8-1, Saiwai-cho, Fuchu, Tokyo, 183-0054, Japan}
\begin{abstract}
In an effort to measure the Rossiter-McLaughlin effect for the
TRAPPIST-1 system, we performed high-resolution spectroscopy during
transits of planets e, f, and b. The spectra were obtained with the
InfraRed Doppler spectrograph on the Subaru 8.2-m telescope, and
were supplemented with simultaneous photometry obtained with a 1-m telescope of the Las Cumbres Observatory Global Telescope.  By analyzing the anomalous radial
velocities, we found the projected stellar obliquity to be 
$\lambda=1\pm 28$ degrees under the assumption that the 
three planets have coplanar orbits, although we caution that 
the radial-velocity data show correlated noise of unknown origin.  We
also sought evidence for the expected deformations of the stellar
absorption lines, and thereby
detected the ``Doppler shadow'' of planet b with
a false alarm probability of $1.7\,\%$. 
The joint analysis of the observed residual cross-correlation map 
including the three transits gave $\lambda=19_{-15}^{+13}$ degrees. 
These results indicate that the the TRAPPIST-1 star is not
strongly misaligned with the common orbital plane of the planets,
although further observations are encouraged to verify this
conclusion.
\end{abstract}
\keywords{
planets and satellites: individual (TRAPPIST-1) --
techniques: photometric --
techniques: radial velocities --
techniques: spectroscopic --
}

\section{Introduction}\label{s:intro}

Measuring the stellar obliquity (the angle between a star's spin
axis and the orbital axis of a planet) is an important method to
probe the dynamical history of exoplanets.  Unlike the Solar System,
in which the planets' orbital planes are aligned with the solar equator to within
about $6^\circ$, measurements of the
Rossiter-McLaughlin (RM) effect \citep{1924ApJ....60...15R,
  1924ApJ....60...22M} have revealed that stars with short-period
giant planets have a broad range of obliquities
\citep{2010A&A...524A..25T, 2012ApJ...757...18A}.
This finding has led to
many proposals for mechanisms that could have tilted the orbital or
the spin axes \citep[see, e.g.,][]{2007ApJ...669.1298F,
  2008ApJ...678..498N, 2010MNRAS.401.1505B, 2012ApJ...758L...6R,
  2013ApJ...778..169B}.

The obliquities of stars with multiple transiting planets in coplanar
orbits --- particularly those lacking massive companions on wide orbits --- are
interesting and relatively unexplored.  Such systems may provide the
opportunity to probe the ``primordial'' obliquities of planet-hosting
stars. The orbital coplanarity suggests that these systems have
not experienced major dynamical rearrangements due to planet-planet
scattering \citep{2012ApJ...761...92F}, and the lack of a massive companion 
on a wide orbit removes the
possibility that the planetary orbital plane has been tilted by secular
gravitational perturbations.  The stellar
obliquity has only been measured for a small number of multiple-planet systems.
In most cases, the 
obliquity is small \citep[e.g.,][]{2012Natur.487..449S, 2012ApJ...759L..36H,
  2013ApJ...771...11A}, although there are at least two systems with
large misalignments
\citep{2013Sci...342..331H, 2019A&A...631A..28D}.

TRAPPIST-1 is a cool M dwarf
($T_{\rm eff} \approx 2{,}550$\,K) known to host seven
transiting planets \citep{2016Natur.533..221G, 2017Natur.542..456G}.
The system has attracted particular attention because the third,
fourth, and fifth planets (e, f, and g) reside inside or near the
``habitable zone" of the host star.  The orbital periods of the planets
are also close to mean-motion resonances, leading to detectable
transit-timing variations (TTVs) that have been used to constrain the
masses of all seven planets.  The latest such analysis found that the
planets have a mainly rocky composition \citep{2018A&A...613A..68G}.
The existence of resonances and the coplanarity of the orbits both suggest
that the planets were drawn together by convergent migration within
the protoplanetary disk \citep[i.e., type I
  migration;][]{2010exop.book..347L}.  Moreover, no stellar companion
has been detected which could have disturbed the initial alignment
between the stellar equator and the protoplanetary disk
\citep{2016ApJ...829L...2H}.

During a 6-hour interval on the night of UT 2018 August 31, three of the
planets in the TRAPPIST-1 system (planets e, f, and b) sequentially
transited the star. We took advantage of this opportunity to try and measure
the stellar obliquity, by performing high-resolution infrared
spectroscopy with the InfraRed Doppler (IRD) instrument
mounted on the Subaru 8.2-m
telescope on Maunakea \citep{2018SPIE10702E..11K}.
This new instrument has a resolution of approximately
70{,}000 and a spectral range of 0.95 to 1.75 $\mu$m; TRAPPIST-1
is extremely faint in the visible, and these kinds of characterizations
are only feasible in the near infrared. 

\section{Observations}\label{s:observation}

\subsection{High Dispersion Spectroscopy with Subaru/IRD}\label{s:IRD}

We observed TRAPPIST-1 with the IRD spectrograph
for 7 hours spanning almost all of
the transit of planet e, followed by the complete transits of planets
f and b.\footnote{$\mathrm{BJD}=2458361.8$~to~$2458362.1$}  The exposure
time was 300 seconds.  Beforehand, we observed a rapidly rotating A0
star \citep[HD 195689;][]{1993yCat.3135....0C} to help with telluric
corrections.  For simultaneous
wavelength calibration, the light from
a laser-frequency comb (LFC) was injected
into the spectrograph using a second fiber. To support our analysis, we also
used a few IRD spectra of TRAPPIST-1 that were obtained when no
planets were transiting, in 2018 December, 2019 January, and 2019 July.
These spectra were used to construct a telluric-free stellar template
for the RV analysis, exploiting the variation in the stellar RV due to
Earth's orbital motion.

We reduced the raw IRD data with the echelle package
of \texttt{IRAF}, for the most part.
We extracted wavelength-calibrated one-dimensional
(1-d) spectra for each stellar spectrum and the accompanying LFC
spectrum.  On the transit night,
the signal-to-noise ratio (S/N) of the 1-d stellar spectrum
varied within the range of
15 to 25 per pixel at $1\,\mu$m, and 45 to 50 per pixel at $1.6\,\mu$m.


\subsection{Simultaneous Photometry}\label{s:mcdonald}

We also performed time-series photometry of TRAPPIST-1 on the transit
night, using one of the McDonald 1-m telescopes of the Las
Cumbres Observatory Global Telescope (LCOGT) network.  
The main purpose was to provide
stronger constraints on the mid-transit times.  We observed with the
SINISTRO CCD camera and a Cousins $I$-band filter for 4.8 hours
spanning the transits of planets e and
f.\footnote{$\mathrm{BJD}=2458361.72$~to~$2458361.92$} Given the
apparent magnitude of TRAPPIST-1 ($m_I=14$) we used 60-second
exposures.

The images were reduced by the BANZAI pipeline
\citep{2018zndo...1257560M}, and aperture photometry was performed
with a custom code \citep{2011PASJ...63..287F}.  Photometry was also
performed on five nearby and
isolated stars, to allow for correction of the
effects due to differential extinction by the Earth's atmosphere.
The comparison stars were brighter than TRAPPIST-1 in the $I$-band by
1.1 to 1.5 magnitudes.  After trying various sizes for the photometric
aperture, we settled on a radius of 8 pixels, which was found to
minimize the scatter in the differential
light curve.  The bottom panel of Figure
\ref{fig:rv} shows the light curve.

\section{Data Analysis}\label{s:analysis}

\subsection{Radial Velocities}\label{s:RV}

We extracted RVs from the IRD spectra using the following
procedure.\footnote{These procedures will be
described in more detail in a forthcoming paper (Hirano et al., in preparation).}  
The total spectral range was divided into $\approx 1200$ spectral segments, 
each spanning a wavelength range of $\Delta\lambda=0.7-1.0$ nm.
The LFC spectra were used to model
the instantaneous instrumental profile (IP) of each spectral segment.
Using this IP model, we extracted a template spectrum by deconvolving
each TRAPPIST-1 spectrum and removing telluric absorption features. 
The telluric lines are identified and removed using normalized spectra of 
the rapidly-rotating A0 star for most cases, but we also removed those 
lines by fitting a theoretical transmittance \citep{2005JQSRT..91..233C}
in case that no telluric standard star is observed immediately before 
or after our observation of TRAPPIST-1 (UT 2018 August 7 and December 25). 
We shifted the resulting spectra with wavelength into a common frame of reference
by cross-correlating telluric-free spectral segments against a theoretical model spectrum for TRAPPIST-1 \citep{2013MSAIS..24..128A}. 
Then, we median-combined the
spectra to obtain a deconvolved and telluric-corrected template
spectrum, representing our best estimate of the intrinsic spectrum of TRAPPIST-1.

\begin{figure}
\centering
\includegraphics[width=8.5cm]{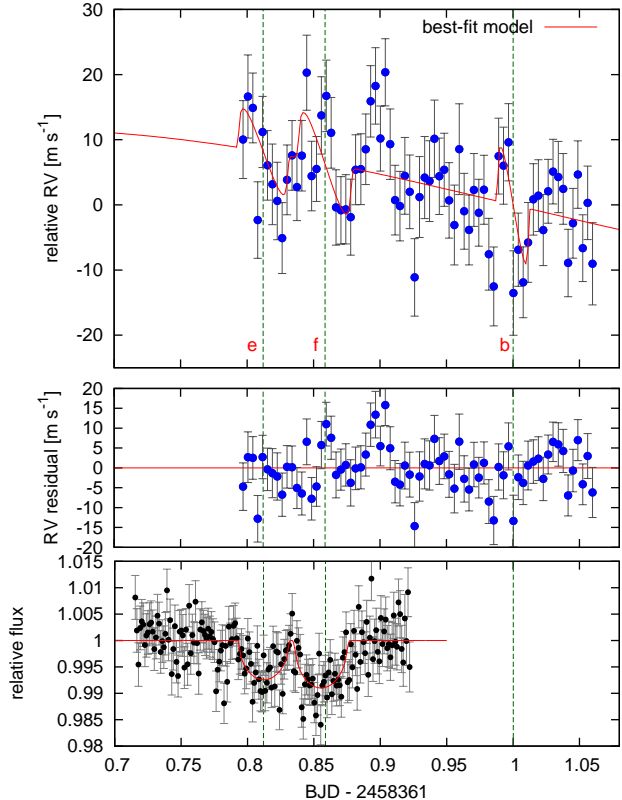}
\caption{
{\it Top:} Relative RVs measured with Subaru/IRD (blue points) and the best-fitting model (red line). Note that the RV value at $T_c$ of planet b is set to zero (the systemic velocity of TRAPPIST-1 ($\approx -52.5$ km s$^{-1}$) was subtracted). 
{\it Middle:} Residuals.
{\it Bottom:} Simultaneous photometry from LCOGT/SINISTRO (black points) and the best-fitting model (red line). The vertical dashed lines are the best estimates of the transit midpoints.
}
\label{fig:rv}
\end{figure}

Using this template spectrum $S(\lambda)$, we modeled each IRD
spectral segment $f_\mathrm{obs} (\lambda)$ as
\begin{eqnarray}
f_\mathrm{obs} (\lambda) = k(\lambda) \times\left[S\left(\lambda\sqrt{\frac{1+v/c}{1-v/c}}\right)\times T(\lambda) \right] \otimes \mathrm{IP},
\end{eqnarray}
where $v$ is the RV, $\otimes$ is the convolution operator, $k(\lambda)$ is
a quadratic function of vacuum wavelength $\lambda$ that was included
to account for the overall continuum variation, $T(\lambda)$ is a
model of telluric transmittance \citep{2005JQSRT..91..233C}, and IP is
the estimated instrumental profile.  After optimizing the model by Markov Chain
Monte Carlo (MCMC) samplings using our customized code for each segment, we
determined the RV value and its uncertainty based on the weighted mean
of $v$ across all the spectral segments that did not show especially
strong telluric absorption features.  The RVs on the transit night are shown
by the blue points in the top panel of Figure~\ref{fig:rv}.

\subsection{Refining Transit Ephemerides}\label{s:timing}

To pin down the mid-transit times, we jointly analyzed the
LCOGT data and the available {\it Kepler} data.  The
{\it Kepler} telescope observed TRAPPIST-1 in short-cadence mode
(one-minute integrations) during Campaign 19  of the {\it K2}
mission.\footnote{$\mathrm{BJD}=2458361.1$ to $2458387.5$}  We downloaded
the pixel-level data from the Mikulski Archive for Space Telescopes
(MAST). The data from the first 8.5 days of the
campaign\footnote{cadence numbers 5008450 to 5020749} were unusable
because of an error in spacecraft pointing.  The {\it K2} data
span ten transits of planet b, and three transits each of planets e
and f.  We performed aperture photometry and reduced the systematic
effects from the telescope's rolling motion following a standard procedure
described by \citet{2017AJ....153...40D}.
We found that a circular aperture with a radius of 2.5 pixels minimized the
out-of-transit scatter in the light curve.

\begin{table}[t]
\centering
\caption{Transit Ephemerides.
These results were used to set Gaussian priors for $T_c$ in the RM analyses.}\label{hyo1}
\begin{tabular}{lcc}
\hline\hline
Planet  & $P$ (days) & $T_c$ ($\mathrm{BJD}_\mathrm{TDB}$) \\\hline
b & $1.51087_{-0.00017}^{+0.00022}$ & $2458362.0021_{-0.0022}^{+0.0016} $ \\
e & $6.09705_{-0.00093}^{+0.00096}$ & $2458361.8113_{-0.0025}^{+0.0028} $ \\
f & $9.2129\pm 0.0010$ & $2458361.8545\pm0.0018$ \\
\hline
\end{tabular}
\end{table}

Although the TRAPPIST-1 planets are known to experience TTVs
\citep{2016Natur.533..221G}, we assumed the periods to be constant
while fitting the LCOGT and {\it K2} data.  This is because the LCOGT and
{\it K2} observations were only separated by about seven days, and the
TTVs on such short timescales are negligible
\citep{2018A&A...613A..68G}.  We performed the transit modeling with
the {\tt Batman} package \citep{2015PASP..127.1161K}, adopting a
quadratic limb-darkening law with freely adjustable coefficients for
each of the two bandpasses.  The other transit parameters (radius
ratio $R_p/R_s$, impact parameter $b$, period $P$, and reference
mid-transit time $T_c$) were assumed to be the same across both data
sets.  We assumed the orbits to be circular, and imposed a Gaussian
prior on the mean stellar density of $50.7^{+1.2}_{-2.2}\,\rho_\odot$ \citep{2016Natur.533..221G},
which acts effectively as a constraint on the scaled semi-major axis
$a/R_s$.  We performed an MCMC analysis with
the {\tt emcee} package \citep{2013PASP..125..306F}.  Table \ref{hyo1}
gives the results for the orbital periods and transit midpoints.

\subsection{Modeling of the RM Effect}\label{s:RM}

The RVs on the transit night (Figure~\ref{fig:rv}) show some of the
patterns that are expected from the RM effect, along
with some other variations that are not associated with the
transits.  Because of the low S/N ratio and lack of RV points
before the transits of planet e and f, we
decided to assume that the projected obliquity $\lambda$ is the same
for all three planets, instead of trying to determine
the projected obliquity relative to each of the planetary orbits.
To model the anomalous
RV due to the RM effect, we used the
formulas presented by  \citet{2011ApJ...742...69H}.
The free parameters were $\lambda$, the projected rotation velocity
$v\sin i_s$, and the slope and intercept of a linear function of time.
The intercept represents the arbitrary RV offset $\gamma$, and the
slope represents the apparent acceleration of star
throughout the transit night. This is partly from
the gravitational acceleration from the combined pull of all the planets, but
may also include systematic effects from stellar activity, telluric
effects, or instrumental variations. To avoid having to mitigate the
effects of long-term stellar activity and account for the orbital
motion of all seven planets, we only fitted the data from the transit
night. The transit midpoints were allowed to vary subject to 
Gaussian priors based on the values in Table \ref{hyo1}. 
The transit impact parameter $b$ is known to correlate with $v\sin i_s$
and $\lambda$, especially when $b$ is close to zero \citep[e.g.,][]{2011ApJ...738...50A},
and hence we also let $b$ float with Gaussian priors based on the literature values
\citep{2017Natur.542..456G}. 
The other transit parameters were held fixed at the values reported in the
literature \citep{2017Natur.542..456G}.

We performed an MCMC analysis using
the code described by \citet{2016ApJ...825...53H}.
The best-fitting model has $\chi^2=81.4$
with 68 degrees of freedom.  Table~\ref{hyo2} gives the results for
the parameters, which are based on the 16\%, 50\%, and 84\% levels of
the marginalized cumulative posterior distributions.  The red curve in
the top panel of Figure \ref{fig:rv} shows the best-fitting model, and
the middle panel shows the residuals when the model RVs are subtracted
from the observed RVs.

\begin{table*}[tb]
\centering
\caption{Estimated RM Parameters and Their Priors. The symbol $\mathcal{U}$ represents
a uniform prior. }\label{hyo2}
\begin{tabular}{lcccc}
\hline\hline
Planet  & b & e & f & Imposed Prior \\\hline
\textbf{Effective RV analysis} &&& \\
$v\sin i_s$ (km s$^{-1}$) (three transits combined) & & $1.50 \pm 0.37$ & & $\mathcal{U}\,[-\infty, +\infty]$ \\
$\lambda$ (deg.) (three transits combined) & & $1 \pm 28$ & & $\mathcal{U}\,[-180, +180]$ \\\hline
\textbf{Doppler shadow} &&& \\
$v\sin i_s$ (km s$^{-1}$) & $2.16_{-0.55}^{+0.17}$  & $1.74_{-0.75}^{+0.43}$ &  $1.85_{-0.73}^{+0.20}$ & $\mathcal{U}\,[0.6, 2.4]$\\
$\lambda$ (deg.) & $15_{-30}^{+26}$ & $9_{-51}^{+45}$ & $21\pm 32 $ & $\mathcal{U}\,[-180, +180]$\\
$v\sin i_s$ (km s$^{-1}$) (three transits combined) & & $2.04_{-0.16}^{+0.18}$ & & $\mathcal{U}\,[0.6, 2.4]$\\
$\lambda$ (deg.) (three transits combined) & & $19_{-15}^{+13}$ & & $\mathcal{U}\,[-180, +180]$\\
\hline
\end{tabular}
\end{table*}

The result for the projected obliquity is $\lambda=1\pm 28$
degrees, consistent with spin-orbit alignment of the TRAPPIST-1
system.  The result for
the projected rotation velocity is $v\sin i_s=1.49_{-0.37}^{+0.36}$
km\,s$^{-1}$.  As a consistency check, it is useful to compare this
result for $v\sin i_s$ with the value of the rotation velocity,
$v=2\pi R_s/P_{\rm rot}$, based on the stellar radius $R_s$ and
rotation period $P$.  To estimate $P_{\rm rot}$, we inspected the {\it K2} 
light curve for signs of periodic photometric variability.  The
Lomb-Scargle periodogram \citep{1976Ap&SS..39..447L,
  1982ApJ...263..835S} shows a peak at 3.28\,days, and the
auto-correlation function leads to a period of 3.25\,days. Both of
these estimates are in agreement with the period reported by
\citet{2017NatAs...1E.129L}. In combination with $R_s=0.117\,R_\odot$
\citep{2016Natur.533..221G}, 
the calculated rotation velocity is 1.8~km\,s$^{-1}$. This is
consistent with our result for $v\sin i_s$,
and therefore consistent with a low
stellar obliquity.  We note, though, that our result for $v\sin i_s$
is lower than the value of $6\pm2$~km\,s$^{-1}$ reported by
\citet{2010ApJ...710..924R}, which we believe was mistaken.
In fact, the high value of $v\sin i_s$ reported earlier made the
prospect of RM observations appear easier than it was in reality;
for instance, \citet{2016MNRAS.462.4018C} predicted an RM amplitude 
of $40-50$ m s$^{-1}$ based on this larger $v\sin i_s$.
To make sure that our measurement of $\lambda$ is not vulnerable to 
$v\sin i_s$, we fitted the RV data with $v\sin i_s$ being fixed at $1.8$ 
km s$^{-1}$. The result for $\lambda$ was unchanged, as expected 
($\lambda=4_{-32}^{+30}$ degrees).

The residuals exhibit correlated patterns, particularly
between $0.85$ and $0.90$ on the time axis of Figure~\ref{fig:rv}. We
do not know the origin of those features, and can only speculate that
they are from stellar activity, flares, or instrumental effects.
In order to quantify the impact of this correlated noise on $\lambda$, 
we performed a numerical experiment using the ``prayer bead" method 
\citep[e.g.,][]{2009ApJ...699..478D} as follows;
Assuming that the residual data between the observed RVs and the best-fit model
(middle panel of Figure \ref{fig:rv}) reflect the magnitude and timescale 
for the correlated noise, we created a series of mock RV data sets by 
adding cyclically permuted residual RVs (time stamps shifted one by one) 
to the best-fit model in Figure \ref{fig:rv}, thus generating $72$ sets of mock RV data. 
We then fitted each mock data set using the same code as above. 
As a result of this numerical experiment, we found that the scatter in the best-fit 
$v\sin i_s$ and $\lambda$ were $0.51$ km s$^{-1}$ and $31$ degrees, respectively. 
These estimates of systematic errors are comparable to the statistical errors 
reported in Table \ref{hyo2}.

\section{The Doppler shadow}\label{s:tomography}

Given the large statistical and systematic uncertainties in $v\sin i_s$
and $\lambda$, it is reasonable to question the result for our fit to 
IRD RV data. We were thereby
motivated to perform a second analysis of the RM effect, based on
modeling the distortions in the line profiles rather than extracting
an effective RV \citep[see, e.g.,][]{2007A&A...474..565A, 2010ApJ...709..458H,
  2010MNRAS.407..507C}. This technique is often called ``Doppler
tomography,'' although we prefer to refer to the ``Doppler transit''
or ``Doppler shadow'' of the planet, because there is not much
resemblance to the standard meaning of {\it
  tomography}.\footnote{Tomography:~constructing a 3-d image by
  combining a series of 2-d sections or 2-d integrated images obtained
  from different angles, using penetrating radiation.}

In previous analyses of Doppler transits, the spectral line profiles
have been modeled with least-squares deconvolution
\citep[LSD;][]{1997MNRAS.291..658D}. In this case, though, most of the
spectral information is from complex and blended molecular absorption
lines, making it almost impossible to define the spectral continuum or
analyze individual lines. Instead, we used the classical
cross-correlation function (CCF) technique to extract the mean line
profile. To generate a cross-correlation template, we deconvolved the
individual IRD spectra using the instantaneous IP and a theoretical
rotation/macroturbulence broadening kernel \citep{2005oasp.book.....G}. 
We employed the iterative/recursive
deconvolution technique described by \citet{1994rhis.conf...24C} to
extract the mean deconvolved spectrum of TRAPPIST-1, using
a large number of out-of-transit spectra taken on multiple nights. 
For the deconvolution, we assumed $v\sin i_s=1.8$ km s$^{-1}$ based on 
the rotation period measurements described in
Section \ref{s:RM}, and the macroturbulent velocity of $\zeta=1$ km s$^{-1}$,
following \citet{1998ApJ...498..851V} and \citet{2006ApJ...652.1604B}.
The resulting deconvolved, normalized spectrum,
denoted by $S_\mathrm{intrinsic}$, represents our best estimate of
TRAPPIST-1's spectrum in the absence of rotation.  This is needed to
represent the emergent spectrum from the portion of the star that is
blocked by a transiting planet.

\begin{figure}
\centering
\includegraphics[width=8.5cm]{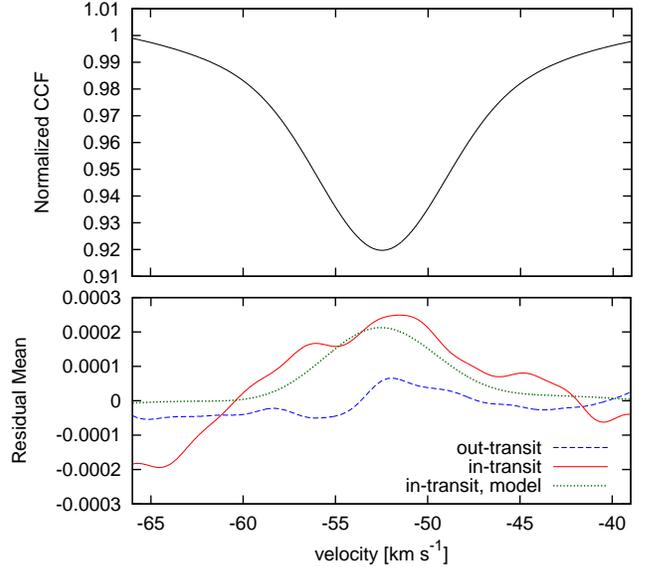}
\caption{
{\it Top:} 
Barycentric velocity (RV) vs.\ mean out-of-transit CCF between observed TRAPPIST-1's spectra and its flipped 
template (i.e., $1-S_\mathrm{intrinsic}$). 
Note that the CCF has a minimum because the features being correlated are absorption lines.
{\it Bottom:} 
Mean residual CCF during the transit of TRAPPIST-1b, stacked at the instantaneous planet shadow position in the
profile (red solid line). Mean out-of-transit residual CCF, covering the same time interval, is shown by the
blue dashed line. The green dotted line plots the theoretical mean in-transit residual CCF model. 
See Section \ref{s:discussion} for more details.  
}
\label{fig:ccf}
\end{figure}

For the CCF analysis, we divided the individual IRD spectra by
the normalized A0 star spectrum to
remove the telluric lines, taking into
account the differences in airmass
between the observation of the standard star and those of TRAPPIST-1
with a plane parallel atmosphere \citep[e.g.,][]{2018PASJ...70...84K}.
Then, we cross-correlated each processed IRD spectrum
with $1-S_\mathrm{intrinsic}$, and normalized the resulting
CCF. Finally, we corrected for
the barycentric motion of the Earth for each CCF and translated the
CCFs to a common velocity scale.  The top panel of Figure
\ref{fig:ccf} shows the mean CCF when
no transits were occurring.

\begin{figure*}
\centering
\includegraphics[width=18cm]{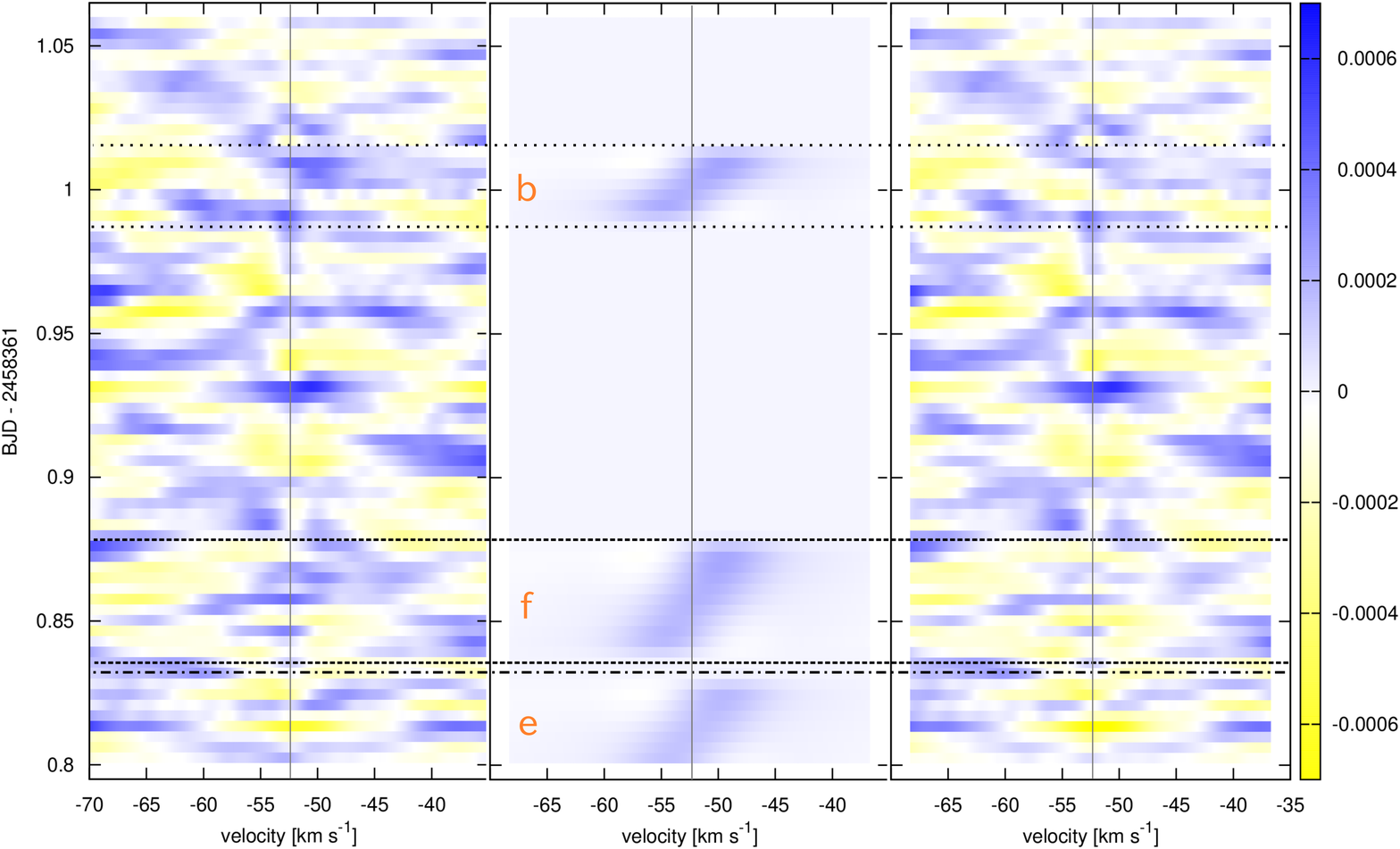}
\caption{
{\it Left:} Observed residual CCF map (velocity-time space) based on the IRD spectra for TRAPPIST-1. 
{\it Middle:} Best-fit theoretical model for the residual CCF map after the MCMC analysis. 
The observed CCF map exhibits some instantaneous variations on a timescale of 10--20 minutes. 
The impact of the correlated noise in discussed in Section \ref{s:discussion}. 
{\it Right:} Observed residual CCF map after subtracting the best-fit model. }
\label{fig:DT}
\end{figure*}

To visualize the mean line-profile variations, we subtracted
the mean out-of-transit CCF from each individual CCF, and examined the
resulting residuals as a function of time. The raw residual
CCF map exhibited an overall low-frequency modulation which is most
likely due to a combination of detector persistence, variations in
the signal-to-noise ratio, and the imperfect removal of telluric
lines.  To account for this slow modulation, we fitted a fourth-order
polynomial function of time to the residual CCF and subtracted the
polynomial from each velocity column. Since the timescale of
  CCF variations due to a transit (30--60 minutes) is much smaller
  than the observing span (7 hours) on that night, this step would not
  significantly diminish the planetary signal. The left panel of
Figure \ref{fig:DT} shows the resulting time series of residual CCFs,
which were searched for the Doppler shadow of each planet. 

The transit ingress and egress times
are indicated by pairs of horizontal lines
(for planet e, f, and b, from bottom to top).  Given the star's rotation
velocity of about 2~km s$^{-1}$, any features on the photosphere
should be shifted by no more than $2$ km s$^{-1}$ from the barycentric RV of
$-52.5$~km\,s$^{-1}$.  If the stellar obliquity is small, the
planetary shadow appears as a loss of
light that moves from the
blue (left) side to the red (right) side of the line profile. In the figure, this would
correspond to a blue bump that progresses from the lower left to the
upper right of the rectangle spanned by the transit. Such a traveling
bump does appear to be barely discernible for TRAPPIST-1b, but is not evident
for the other planets.

For quantitative analysis, we created about 2{,}000 mock spectra mimicking
the IRD data during a transit of each planet, following
\citet{2011ApJ...742...69H}. To create each mock spectrum, we began
with $S_\mathrm{intrinsic}$, and applied the same blaze function, IP,
etc., as the real data. This was done for a range of possible values
of $v\sin i_s$ from 0.6 to 2.4~km\,s$^{-1}$, and for the full range of
positions of the planet on the stellar disk.  We subjected the mock
transit spectra to the same CCF analysis as the real data.  The grid
of values was fine enough to allow the residual CCF to be computed reliably
for any values of $v\sin i_s$ and
planet position, using linear interpolation.

This provided all the necessary ingredients for a model of the
residual CCFs.  We fitted the model to the real data
for each of the three transits (i.e., for each planet). 
We performed an MCMC analysis to determine the parameters.  
We adopted uniform priors on $v\sin i_s$ and $\lambda$
and imposed the same Gaussian priors on the midtransit times 
and impact parameters $b$ that were
described earlier.  The only freely adjustable parameters were $v\sin
i_s$ and $\lambda$ for each transit; the other transit parameters were held fixed at
the values reported by \citet{2017Natur.542..456G}.  Table \ref{hyo2}
gives the results.  
The best-fit obliquity $\lambda$ has large uncertainties for
individual transits, but all the fitting results prefer a low obliquity 
(prograde orbit). Motivated by this fact, we performed a joint
analysis assuming the same values of $v\sin i_s$ and $\lambda$ for all 
the three transits, and fitted the whole residual CCF data on the night. 
The projected obliquity was found to be $\lambda=19_{-15}^{+13} $ degrees, 
consistent with the result for the RV fit. 
The best-fit model of the residual CCF time series and the observed
CCF map with the best-fit model subtracted are depicted in the middle 
and right panels of Figure \ref{fig:DT}, respectively.

\section{Discussion and Summary}\label{s:discussion}

Besides $\lambda$, another parameter that deserves discussion is the
overall slope of the RV time series on the transit night.  If this
apparent acceleration were entirely due the gravitational forces from
the planets, the dominant effect would be from planet b.  Under this
interpretation, the best-fitting slope corresponds to an RV
semi-amplitude for planet b of $K=11.6_{-2.4}^{+2.5}$ m s$^{-1}$.  
Alternatively, we could fit the RV data with the 
$\dot{\gamma}$ parameter, which was found to be $-36.9\pm 9.5$ m s$^{-1}$ day$^{-1}$.
This is how the result is reported in Table~\ref{hyo2}.  The value of $K$ is
significantly larger than the expected value of $\approx\,3$~m\,s$^{-1}$
based on the TTV-derived planet mass \citep{2018A&A...613A..68G}.
Although it is possible that the planet mass was underestimated, or
that there is another massive planet lurking undetected in the system,
it seems more likely that the observed RV slope is produced mainly by
systematic effects due to stellar
activity, imperfect removal of telluric lines,
or instrumental effects.
Instrumental systematics can result from the
persistence of a strong signal in the detector. Immediately before
observing TRAPPIST-1, we observed a very bright rapid rotator as a
telluric standard; it is possible that the signal from this star
persisted during some of the early TRAPPIST-1 spectra during the
transits of planets e and f. Furthermore, our RV pipeline cannot identify 
and model telluric lines perfectly, leading
to systematic errors in the stellar template. Such errors have been
shown capable of producing a spurious drift in the apparent RVs,
similar to the one seen here (Hirano et al., in preparation).

To check the impact of this type of systematic RV on the inferred value of $\lambda$, we
reanalyzed the RV data with a model consisting of circular orbits for
all seven planets.  We adopted the expected values for the RV
amplitudes from \citep{2018A&A...613A..68G}.  The amplitudes of
all the planetary signals were held fixed at their expected values,
except for planet b, for which we imposed a Gaussian prior. (The
other planets lead to only a small variation in the calculated RV, on
the order of 1~m\,s$^{-1}$.) The fit was not as good, and the
resulting RM parameters were $v\sin
i_s=1.88_{-0.43}^{+0.42}$~km\,s$^{-1}$ and $\lambda=-39_{-15}^{+20}$
degrees.  These results are still consistent with spin-orbit alignment
within $2\,\sigma$, but in this case we cannot rule out a moderate
spin-orbit misalignment.  Further observations of the system are
needed to understand the reason for the aberrant RV slope.

One might also wonder about the statistical significance of the
detection of the Doppler shadow in the residual CCFs. To investigate this issue, we stacked the residual CCFs obtained during the transit
of planet b, after Doppler-shifting (aligning) each residual CCF so
that the theoretical position of the shadow is always at the same
velocity ($\approx\,-52.5$~km\,s$^{-1}$). The red curve in the bottom
panel of Figure~\ref{fig:ccf} shows the resulting stacked in-transit
residual CCF.  The green curve in the same panel shows the in-transit
residual CCF calculated using the parameters
of the best-fitting model.  The blue curve
is a stacked out-of-transit residual CCF based on approximately the
same amount of data as the in-transit version. The Doppler shadow is
detected to the extent that the peak in the red curve exceeds the
level of the blue curve.

We used Monte Carlo simulations to estimate the false alarm
probability (FAP) that the feature in the stacked in-transit residual
CCF was produced by chance.  First, we split the observed
out-of-transit residual CCFs into 17 segments, each having a few frames
(so that the data kept correlated noise of this timescale). 
We then resampled the segments to create a mock time series of residual CCFs in a random order.  No planetary signal was injected.  We then fitted the
mock data with the same code that was used on the real data, assuming
that there is a planet signal. In this way, we produced a mock version
of the mean in-transit residual CCF.  We repeated the mock-data
analysis for 1{,}000 trials, recording the peak value of the mean
in-transit residual CCF.  We found that the peak value in the fake
data exceeds the peak value in the real data in 1.7\% of the trials,
suggesting that it is unlikely that the observed CCF bump was produced
by chance.

Based on all these tests, we have a reasonable level of confidence
that the Doppler shadow of planet b was detected and is consistent
with a low stellar obliquity. It is also reassuring that the analysis
of the pattern of anomalous RVs led to the same conclusion.  On the
other hand, similar mock-data analyses for TRAPPIST-1e and f resulted
in FAPs of 36\% and 22\%, respectively, implying that we cannot have
any confidence in the detection of the Doppler shadows of those
planets.

Our result supports the idea that the known planets in the TRAPPIST-1
system achieved their compact configuration through convergent
migration, and did not experience any substantial misaligning torques
from processes such as planet-planet scatterings or long-term
gravitational perturbations from a massive outer companion on an
inclined orbit.  It is unlikely that any primordial obliquity has been erased by tidal realignment between the star and these low-mass planets.  
An order-of-magnitude estimate for the tidal realignment timescale for planet b, 
calculated as in \citet{2011A&A...535A..94B} using the dissipation of 
\cite{2010ApJ...723..285H}, ranges from $\sim 10^{11}$ years 
(assuming $R_\star\approx 0.5\,R_\odot$ in the past) to
$\sim 10^{17}$ years (adopting the current stellar radius).

Our analysis was conducted mostly under the assumption that the orbits of the
TRAPPIST-1 planets are coplanar.  We were unable to test this
assumption by measuring the mutual inclinations between the planets.
The mutual inclinations might be measurable in the future using
repeated observations of Doppler transits to give a higher
signal-to-noise ratio.  The mutual inclination between two planetary
orbits might also be measured by observing the photometric effect of a
planet-planet eclipse during a double transit event
\citep{2012ApJ...759L..36H}.  This would be another important clue to
understand the architecture and dynamical history of the TRAPPIST-1
system.

Despite the limitations of the data, our observation of the Doppler
transits in the TRAPPIST-1 system are the first such observations, to
our knowledge, for such a low-mass star.  No other results have been
reported for stars cooler than 3500\,K.  By performing additional
observations with the IRD and other new high-resolution infrared
spectrographs, a new window will be opened into the orbital
architectures of planetary systems around low-mass stars.

\acknowledgments
This work is based in part on data collected at Subaru Telescope, which is operated by the National Astronomical Observatory of Japan, 
and makes use of observations from the LCOGT network.
The data analysis was carried out, in part, on the Multi-wavelength
Data Analysis System operated by the Astronomy Data Center (ADC),
National Astronomical Observatory of Japan.  This work is supported
by JSPS KAKENHI Grant Numbers 16K17660, 19K14783, 18H05442, 15H02063, 
and 22000005, and by the Astrobiology Center Program of
National Institutes of Natural Sciences (NINS) (Grant Number AB311017). 
JNW thanks the Heising-Simons foundation for support.  
LMW is supported by the Beatrice Watson Parrent Fellowship. 
SA and MH acknowledge support from the Danish Council for Independent Research through the DFF Sapere Aude Starting Grant No. 4181-00487B, and the Stellar Astrophysics Centre which funding is provided by The Danish National Research Foundation (Grant agreement no.: DNRF106).
This work has been carried out within the framework of the NCCR PlanetS supported by the Swiss National Science Foundation.
\software{{\tt batman} \citep{2015PASP..127.1161K}, 
{\tt emcee} \citep{2013PASP..125..306F}, 
{\tt IRAF} \citep{1986SPIE..627..733T, 1993ASPC...52..173T}
}



\end{document}